\documentstyle[twoside,fleqn,espcrc2]{article}
\input{psfig}
\def\bold#1{\setbox0=\hbox{$#1$}%
     \kern-.025em\copy0\kern-\wd0
     \kern.05em\copy0\kern-\wd0
     \kern-.025em\raise.0433em\box0 }
\def\slash#1{\setbox0=\hbox{$#1$}#1\hskip-\wd0\dimen0=5pt\advance
       \dimen0 by-\ht0\advance\dimen0 by\dp0\lower0.5\dimen0\hbox
         to\wd0{\hss\sl/\/\hss}}
\def\lq{\left [}
\def\rq{\right ]}

\newcommand{\be}{\begin{equation}}
\newcommand{\ee}{\end{equation}}
\newcommand{\bea}{\begin{eqnarray}}
\newcommand{\eea}{\end{eqnarray}}
\newcommand{\nn}{\nonumber}
\newcommand{\dd}{\displaystyle}

\newcommand{\qq}{<0|{\bar q} q|0>}

\newcommand{\AmS}{{\protect\the\textfont2
  A\kern-.1667em\lower.5ex\hbox{M}\kern-.125emS}}

\title{On quark-hadron duality in the heavy quark sector}

\author{P. Colangelo
\address{Istituto Nazionale di Fisica Nucleare - Sezione di Bari \\ 
via Amendola n.173, I-70126 Bari, Italy}%
        \thanks{Talk given at the High Energy Europhysics Conference
QCD 97, Montpellier 3-9 July 1997}
}
       
\begin{document}

\begin{abstract}
I discuss possible failures of local quark-hadron duality
in the system of heavy-light quark mesons.
As an example, I consider a correlator of two currents comprising heavy quark 
operators, and I compare 
the OPE expression with the result obtained by a complete insertion of 
hadronic states.
After a smearing procedure, OPE and hadronic spectral functions agree with 
each other. However, the local behaviour of the two functions is different, and 
the difference manifests itself in a term which is absent in the OPE.
\end{abstract}

\maketitle

\section{Introduction}

Quark-gluon/hadron duality represents a basic concept in the theoretical 
description of inclusive hadronic processes.
In general, it means that some hadronic rates can be computed as the 
corresponding partonic rates, and that
high energy processes 
can be computed in terms of hadronic matrix elements of
operators in an expansion, the Operator Product Expansion (OPE), whose
leading term is represented by the perturbative QCD expression. It is duality
which allows us to extrapolate
from the deep Euclidean region, where the OPE is defined, to the 
Minkowski domain, where physical observables are measured. 
\par
An OPE-based approach has been recently proposed for the calculation of 
semileptonic and nonleptonic decay rates of hadrons 
containing one heavy quark \cite{c}. The basic idea consists in the
expansion in inverse powers of the heavy quark mass $m_Q$.
In this approach, however, a distinction must 
be maintained between semileptonic and nonleptonic decays. 
In general, hadronic and OPE amplitudes cannot 
be identical even at very high momentum transfer, 
due to the different production thresholds of
multiparticles and of quarks and gluons. In the case of semileptonic 
heavy hadron decays this difficulty can be avoided, since 
one has to integrate over lepton variables, which amounts 
to a smearing of the OPE width. The equality between smeared OPE
and hadronic widths is referred to as
{\it global duality} (different 
from {\it local duality}, i.e. without smearing). It is generally 
believed that global duality holds between quark-gluon and hadronic cross 
sections \cite{wein}; for $B$ and $\Lambda_b$ semileptonic decays global 
duality has been extensively analyzed in  \cite{b,sv}.
\par
For nonleptonic heavy hadron decays there are
no lepton momenta to be integrated; therefore, one cannot invoke
global duality to prove the identification of OPE and hadronic observables,
and therefore local duality must be assumed. 
In this case, the validity of the OPE-based $1/m_Q$ expansion
for the computation of
nonleptonic heavy hadron inclusive decay rates appears to be more debatable. 
Indeed, in \cite{a} it was suggested that the discrepancy between
the prediction $\displaystyle{\frac{\tau(\Lambda_b)}{\tau(B_d)}>0.9}$
\cite{neubert}
and the experimental result
$\displaystyle{\frac{\tau(\Lambda_b)}{\tau(B_d)}=0.78\pm0.04}$ \cite{richman}
might be solved assuming a violation of local quark-hadron duality, with the
appearance of a ${\cal O}(\frac{1}{m_Q})$ correction 
not predicted by OPE.

At the moment, understanding 
the origin of this possible correction to the
OPE is a difficult task.
However, it is interesting to investigate possible 
violations of local duality in some definite model. 
\par
Several recent studies can be found in the literature
concerning the validity of the quark-gluon/hadron duality in 
connection with non-perturbative QCD applications
in the heavy quark sector \cite{shif1}-\cite{grinstein}. 
In particular, in \cite{shif1} the 
two-point function involving the difference between scalar and pseudoscalar
heavy-light currents was considered as a benchmark to  investigate
global aa well as local duality properties. 
In the chiral limit (for the light flavour) and the
infinite mass limit (for the heavy flavour), the coefficients of the OPE
for the correlator can be calculated analytically. Then, the spectral function
derived form a particular hadronic model
(in the time-like region) is expanded in a power series 
(in the space-like region) and compared with the {\it exact} result. From
this comparison one can gauge the validity of (local) duality.

Here, I want to discuss how the results are stable against 
modifications of the hadronic model. 
Evaluating the two-point function by a complete insertion of
states derived from a relativistic constituent quark model, I'll show
that an explicit violation of local duality can be detected, in the form
of an unexpected correction to the OPE \cite{col97}.

\section{Example of violation of local duality}

I consider the correlator \cite{shif1} 
\bea
&&\Pi(q)=\frac{i}{4} \int dx e^{i q x} 
[<0|T(J_{S}(x)J_{S}^\dagger(0))|0> - \hfill\nn\\
&&<0|T(J_{P}(x)J_{P}^\dagger(0))|0>]=\Pi_S(q)-\Pi_P(q),
\label{1}\eea
with 
$J_S(x)={\bar Q}(x) q(x)$,
$J_P(x)={\bar Q}(x) i \gamma_5 q(x)$, 
$Q(x)$ and $q(x)$ being heavy and light quark operators, respectively. 
In the chiral limit, $m_q \to 0$, $\Pi(q)$ vanishes in perturbation theory. 
In the infinite heavy quark mass limit, $m_Q \to \infty$, it is convenient
to choose 
$q^\mu~=~(m_Q - \epsilon, {\vec 0})$, so that
the correlator becomes a function of $\epsilon$:
\be
\Pi(\epsilon)~=~\frac{1}{4} \int_0^{ + \infty} d \tau 
e^{- \epsilon \tau} \Phi (\tau)~~~~~~~~~(\epsilon>0) \;\;, \label{5}
\ee
$\Phi(\tau)$ representing the non-local quark condensate
\be
\Phi(\tau)~=~<0|{\bar q}(\tau) e^{i g_s\int_0^\tau dy^\mu A_\mu(y)} q(0)|0>~. 
\label{6}
\ee
When $\epsilon >> \Lambda_{QCD}$, the OPE expression for
$\Pi(\epsilon)$ 
\bea
\Pi_{OPE}(\epsilon) ~&=&~\frac{\qq}{4 \epsilon} \times \hfill \nonumber \\
&[& 1 - 
\frac{m_0^2}{8 \epsilon^2} + c_4 \frac{m_0^4}{\epsilon^4}
- c_6 \frac{m_0^6}{\epsilon^6}+...]  \label{7}
\eea
is given in terms of the quark condensate $\qq=(-240 \; MeV)^3$ and
of the mixed quark-gluon condensate, parametrized by
${\dd{m_0^2=\frac{<0|g_s{\bar q} 
\sigma_{\mu\nu} G^{\mu\nu}  q|0>}{\qq}~=~0.8 \pm 0.2\; GeV^2}}$. 
The positive coefficients $c_{2 n}$  depend on the actual form of 
the non-local condensate; notice explicitly the alternating signs in 
Eq.(\ref{7})  and the absence of even powers of $\epsilon^{-1}$.

It is possible to identify a model for $\Pi(\epsilon)$ having 
the expansion (\ref{7}) \cite{shif1}:
\be
\Pi(\epsilon) = \frac{\qq}{4 \bar{\Lambda}}
\beta (\frac{\epsilon + \bar{\Lambda}}{2 \bar{\Lambda}}) \label{8}
\ee
with $\bar \Lambda$ a parameter and
\be
\beta(z) = \frac{1}{2} \; \left[ \psi \left( \frac{z+1}{2} \right) -
\psi \left( \frac{z}{2} \right) \right] \;\;, \label{9}
\ee
$\psi(z)$ being the logarithmic derivative of the Gamma function. 
$\Pi(\epsilon)$ in (\ref{8})
can be written as
\be
\Pi (\epsilon) = \frac{\qq}{2 \bar{\Lambda}}\sum_{j=0}^{\infty} \;
\frac{(-1)^{j}}{\epsilon/{\bar \Lambda}  + 2j + 1}  \label{10}
\ee
and admits the asymptotic expansion
\be
 \Pi (\epsilon)  \;
\sim_{\hskip -0.5cm \vspace*{0.75cm}_{\epsilon\to + \infty}} \; 
\frac{\qq}{4 \epsilon} \;
\sum_{n=0}^{\infty} \; E_{2n}\frac{\bar{\Lambda}^{2n}}{\epsilon^{2n}} 
\label{12}
\ee
($E_{2n}$ Euler numbers).
Comparison of (\ref{12}) with (\ref{7}) indicates that the model 
is able to reproduce
the right power structure of $\Pi(\epsilon)$. The spectral
density associated to (\ref{8}) 
displays an infinite number of equally spaced poles located
along the negative $\epsilon$ axis, with residues having alternating
signs.

The correspondence between (\ref{12}) and (\ref{7}), however, follows from 
the assumed form (\ref{8}) for the correlator. It would be interesting 
to investigate the properties of global and local duality
using  a representation of the correlator obtained from a hadronic 
model more closely related to QCD.
  
Let us compute (\ref{1}) 
by a complete insertion of hadronic states.
We consider
the $N_c \to \infty$ limit, since in this case  
the only contributions are $J^P=0^+$ and $0^-$
single particle states,
contributing respectively
to $\Pi_S$ and $\Pi_P$. We
denote by $|S_n>$ and $|P_n>$  the states, with masses
$M_{S_n}$ and $M_{P_n}$  respectively, and we define the current-particle
matrix elements:
$<0|J_S|S_n>=\frac{M^2_{S_n}}{m_Q} f_{S_n}$,
$<0|J_P|P_n>=\frac{M^2_{P_n}}{m_Q} f_{P_n}$. 
In the limit $m_Q \to \infty$ the masses can be written as:
\be
M_{S_n(P_n)}=m_Q+\delta_{S_n(P_n)} + {\cal O}(\frac{1}{m_Q}) \;\;.
\ee
The binding energies  $\delta_{S_n(P_n)}$ can be obtained by 
solving the relativistic wave equation
\bea
&&[\sqrt{-\nabla^2+m^2_Q} \hfill \nonumber \\
&&+\sqrt{-\nabla^2+m^2_q} + V({\vec r})] \Psi_n({\vec r})=
M_n \Psi_n({\vec r}) \label{releq}
\eea
where the (central) potential $V(r)$ is chosen as $V(r)=\mu^2 r$,
with constant $\mu$ (string tension). 
Masses, wavefunctions and current-particle couplings of 
scalar and pseudoscalar states can be computed applying
WKB methods to (\ref{releq}) \cite{piet,prep}. One gets: 
\bea
\delta_{\ell}^{(n)}&=&\mu {\sqrt{\pi(2 n + \ell + \frac{3}{2}) }} \\
f_{\ell}^{(n)}&=&\sqrt{ \frac{ 3 m_Q \delta_{\ell}^{(n)} } {\pi} } 
\frac{\mu}{M_{n,{\ell}}}~~. \label{28}
\eea
\noindent
Assuming the WKB approximation for the lowest-lying state, an estimate 
can be derived for the string tension:
for $m_b=4.6-4.7$ GeV one has $\mu\simeq 300$ MeV.

The choice of considering 
only single particle states between the currents in the correlator
is not too restrictive, as it can be shown by considering 
the imaginary part of the correlator of scalar currents $\Pi_S$ ($\Pi_P$) in
Eq.(\ref{1}). Computing 
the imaginary part of the quark loop diagram  one has,
for $E=- \epsilon>\to +\infty$:
\be
Im \Pi_S^{OPE}(E) ~\to ~\frac{3 E^2}{8 \pi}~. \label{31} 
\ee
On the other hand, the resonance model gives:
\be
Im\Pi_S^{had}
= \frac{3 \mu^2 E}{8}\sum_{n=0}^\infty \delta \lq E-\mu\sqrt{2 \pi(n+7/4)}\rq.
\label{32}\ee
Although the two expressions (\ref{31}) and (\ref{32}) 
look very different,
after a smearing procedure
\cite{wein} obtained by $\sum_n \to \int dn$,
one has
\bea
Im~\Pi_S^{had}&=& \frac{3 \mu^2 E}{8}\int dn
\delta \lq E-\mu \sqrt{2 \pi(n+7/4)} \rq \hfill \nn \\
&=&\frac{3 E^2}{8 \pi}=Im\Pi_S^{OPE}(E)~. 
\eea
Therefore $\Pi_S(E)$  and  $\Pi_P(E)$  satisfy global duality, at 
least at the leading order for $E\to\infty$.

Let us now compute the correlator (\ref{1}) by the hadronic insertion. We get
\be
\Pi^{had}(\epsilon)=
\frac{3 \mu^3}{8 \sqrt{\pi}}\sum_{n=0}^\infty 
\frac{ (-1)^{n+1}\sqrt{n+3/2}}{\epsilon+\mu\sqrt{\pi(n+3/2)}}\label{30}
\ee
and ($E>0$)
\bea
&&R^{had}(E)=Im \Pi^{had}(E) \hfill\nonumber\\
&&=\frac{3 \mu^2 E}{8}\sum_{n=0}^\infty (-1)^n
\delta \lq E- \mu \sqrt{\pi(n+\frac{3}{2})}\rq. \hfill \label{36}
\eea
From the first four terms in (\ref{7}) we have:
\bea
&&R^{OPE}(E)=Im~\Pi^{OPE} (E) \hfill \nn \\
&=&-\frac{\pi}{4}\qq [ \delta(E)-\frac{m_0^2}{16}
\delta^{\prime \prime}(E)\nonumber \\
&+&c_4\frac{m_0^4}{4!} \delta^{(IV)}(E)
-c_6\frac{m_0^6}{6!}\delta^{(VI)}(E)...] \;\;, \label{37}
\eea
with unknown $c_4$, $c_6$ (${\dd c_4=\frac{5}{64}}$, 
${\dd c_6=\frac{61}{512}}$ in the model \cite{shif1}).
Also in this case
the two expressions look very 
different, but they  can be compared after an 
appropriate  smearing; following \cite{wein} we consider the smeared function
\be
{\bar R}^{had}(E,\Delta)=\frac{\Delta}{\pi}
\int dE^\prime~ \frac{ R^{had}(E^\prime)}{(E-E^\prime)^2+\Delta^2} \label{38}
\ee
and a similar expression for ${\bar R}^{OPE}(E,\Delta)$.
For $\Delta\to 0$ on can prove that ${\bar R}(E,\Delta)\to R(E)$. 
Moreover, since
\be
{\bar R }(E,\Delta)=\frac{1}{2i}\lq\Pi(E-i \Delta)-\Pi(E+i\Delta)\rq,
\label{40}
\ee
for $\Delta>>\Lambda_{QCD}$ one is far from physical singularities.
\par
For a large energy $E$ the two expressions 
$\bar R^{OPE}$ and $\bar R^{had}$
should coincide regardless 
of $\Delta$. Such a request implies:
\be
\qq=-3 \frac{0.51}{2}\frac{\mu^3}{\sqrt{\pi}} \;\;,\label{extra}
\ee
e.g. $\mu=317$ MeV, in agreement with the value derived from the spectrum.
For smaller values of $E$, i.e. $1<E<20$  GeV,
the numerical results concerning 
${\bar R}^{had}(E,\Delta)$ and
${\bar R}^{OPE}(E,\Delta)$ 
are reported in Fig.1, where the ratio
$P={{\bar R}^{had}}/{{\bar R}^{OPE}}$ is plotted
as a function of $E$ for several values of $\Delta$. As expected,
the agreement between ${\bar R}^{had}$ and ${\bar R}^{OPE}$ improves
with increasing $\Delta$; in
particular, for $\Delta=3.0$ GeV the difference does not exceed $20\%$. 
For any value of the smearing parameter $\Delta$  the 
ratio $P$ tends to unity for large energies.
\begin{figure}[htb]
\vskip -0.7cm
\psfig{figure=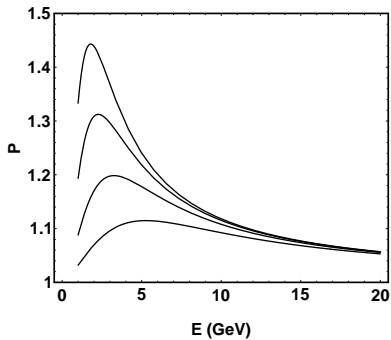,height=5.0 cm}
\vskip -1.cm
\caption{Plot of the ratio $P={{\bar R}^{had}}/{{\bar R}^{OPE}}$.
From top to bottom: $\Delta=1.5,2.0,3.0,5.0$ GeV}
\label{fig:largenenough}
\end{figure}
Therefore, the resonance model satisfies the requirement of 
global duality: the smeared imaginary parts of the correlator, 
when computed by OPE or by hadronic states, agree with each other.

Concerning local duality, 
the expressions are significantly different, since
\be
\Pi^{had}(\epsilon)=-3~\frac{0.51 \mu^3}{8\sqrt{\pi}}\frac{1}{\epsilon}
\lq 1-\frac{{\tilde m}_0}{\epsilon}+0.93\frac{{\tilde m}_0^2}{\epsilon^2}
... \rq~.
\ee
The factor multiplying 
$\epsilon^{-1}$ coincides numerically with $\frac{\qq}{4}$ 
for $\mu=317$ MeV (eq.(\ref{extra})). As for 
the mass parameter ${\tilde m}_0$,
numerically we find ${\tilde m}_0=~560$ MeV. In conclusion one has:
\be
\Pi^{had}(\epsilon)=\frac{\qq}{4\epsilon}
\lq 1-\frac{{\tilde m}_0}{\epsilon}+0.93\frac{{\tilde m}_0^2}{\epsilon^2}
+...\rq ~. \label{last}
\ee
The comparison between 
(\ref{last}) and (\ref{7})  shows a violation of local duality,
which manifests itself in the form of an unexpected term in the 
asymptotic expansion in powers of ${\dd \frac{1}{\epsilon}}$. 
\section{Conclusions}
In the case of a simple correlator of quark currents, 
using a particular quark model and in the $N_c \to \infty$ limit, 
I have shown
an  example of violation of local duality which occurs
in a condition where global duality is verified. 
The difference between the hadronic and the OPE spectral functions is made 
evident by a term which is absent in the expansion predicted by OPE.

Of course, the calculation of the correlators 
required for the evaluation of the $B_d$ and $\Lambda_b$ 
inclusive decay 
widths, is, beyond our present possibilities.
However, the example supports the conjecture 
that the $\Lambda_b$ lifetime problem can be explained  
by the presence of a $1/m_Q$ correction not included in the 
usual OPE expansion.
\vskip 0.5cm
{\noindent \bf Acknowledgments\\} 
\noindent
It is a pleasure to thank C.A.Dominguez and G.Nardulli for their collaboration 
on the topics discussed here. 
I am also grateful to N. Paver and F. De Fazio for interesting discussions.

\vskip 0.1cm
\noindent{\bf Discussions\\}

\noindent {\bf B. Blok} (Technion, Haifa)\\
\noindent {\it How are your results model dependent?\\}

\noindent {\bf P. Colangelo\\}
\noindent {\it 
As discussed, there is a dramatic difference between two choices of the 
hadronic representations of the correlator (\ref{1}). 
Therefore, the results are model dependent, although 
the model I presented seems rather realistic and QCD-inspired. The main point 
is that it allows to detect a mechanism of violation of local quark-hadron 
duality which is similar to what is needed 
to explain the $\Lambda_b$ lifetime problem.}

\vskip 0.5cm
\noindent {\bf M. Neubert} (CERN, Geneva)\\
\noindent 
{\it In my opinion, the violation of global duality found in your model 
($\sim 10-40 \%$) are surprisingly large. Is the E-dependence of these 
violations in accordance of general expectations as discussed, e.g., in the 
previous talk by B.Blok?\\} 

\noindent {\bf P. Colangelo\\}
\noindent {\it 
The numerical details can be ascribed to some features of the model, 
for example to the assumption of vanishing width of the resonances.
What, in my opinion, is instructive is to see how global duality
is recovered, in terms of the energy $E$ and of the 
smearing parameter $\Delta$. }

\end{document}